\documentstyle[twoside,fleqn,espcrc2,psfig]{article}

\newcommand{\AmS}{{\protect\the\textfont2
  A\kern-.1667em\lower.5ex\hbox{M}\kern-.125emS}}
\def\ltsima{$\; \buildrel < \over \sim \;$}
\def\simlt{\lower.5ex\hbox{\ltsima}}
\def\gtsima{$\; \buildrel > \over \sim \;$}
\def\simgt{\lower.5ex\hbox{\gtsima}}

\title{1H0419-577: A TWO-STATE SEYFERT GALAXY?}

\author{M.Guainazzi\address{Beppo-SAX Science Data Center Via Corcolle 19, I-00131 Roma, Italy},%
	A. Comastri, G.Stirpe\address{Osservatorio Astronomico di Bologna, Via Zamboni 33, I-40126, Italy}, 
	W.N.Brandt\address{Dept. of Astr. and Astroph., Penn State University, 525 Davey Lab, University Park, PA 16802, U.S.A.},
	A.Parmar\address{Space Science Department/ESTEC, Noordwijk, The Netherlands},
	E.M.Puchnarewicz\address{Mullard Space Science Laboratory, University College London, Holbury St. Mary, Dorking, Surrey RH5 6NT, U.K.}}
\begin{document}

\begin{abstract}
The preliminary results of the BeppoSAX observation of the radio-quiet
AGN 1H0419-577 are presented. Despite its broad line optical spectrum, the
intermediate X--ray spectrum ({\it i.e.} 2--10~keV)
is flatter than typically observed in Seyfert 1s and no iron line is
significantly detected. Even more intriguingly, a 1992 ROSAT pointed
observation suggests a dramatic ($\Delta \Gamma \simeq 1$) change in
the spectral shape for $E \simlt 2 \ keV$. Such behavior is
briefly discussed in the framework of our current understanding of
Comptonization scenarios in the nuclear regions of radio--quiet AGN.
\end{abstract}

\maketitle

\section{Introduction}

1H0419-577 is a Seyfert 1.5 galaxy, according to Grupe (\cite{gru}). It was
observed by Beppo-SAX within the AO1 program
on September 30 1996, from 06:24:00 UT to 16:40:50 UT, for a total $\simeq 22.6
\ ks$ exposure time in the Medium Energy Concentrator Spectrometer (MECS;
1.8--10.5~keV, \cite{bcc}).
The target was also detected in
the Phoswitch Detector System (PDS, ~\cite{fcd})
at 2.1 $\sigma$ confidence in the 13--36~keV
energy range. The Low Energy Concentrator System (LECS, ~\cite{pmb})
was switched off during the
whole observation due to technical problems.
An optical spectrum was taken at the 1.52~m ESO telescope
on the same night as the Beppo-SAX observation. It displays
``broad'' line components
({\it e.g.} the $<FWHM_{H_{\beta}}> = 1440 \ km \ s^{-1}$,
and H$_{\beta}$ can be split into broad
($FWHM \simeq 3500 \ km \ s^{-1}$) and narrow
($FWHM \simeq 3560 \ km \ s^{-1}$) components.
This confirms the Seyfert 1 nature of this
object.

In this paper errors are quoted at 90\% level of confidence and energies
are in the source rest frame ($q_0 = 0.5$ and $H_0 = 75 \ km \ s^{-1} \
Mpc$ are assumed). Hereafter $\Gamma_{hard}$ ($\Gamma_{soft}$) indicates
the photon index of a power-law above (below) $E \simeq 3 \ keV$.

\section{The X--ray spectrum}

\begin{figure}
\centerline{\psfig{figure=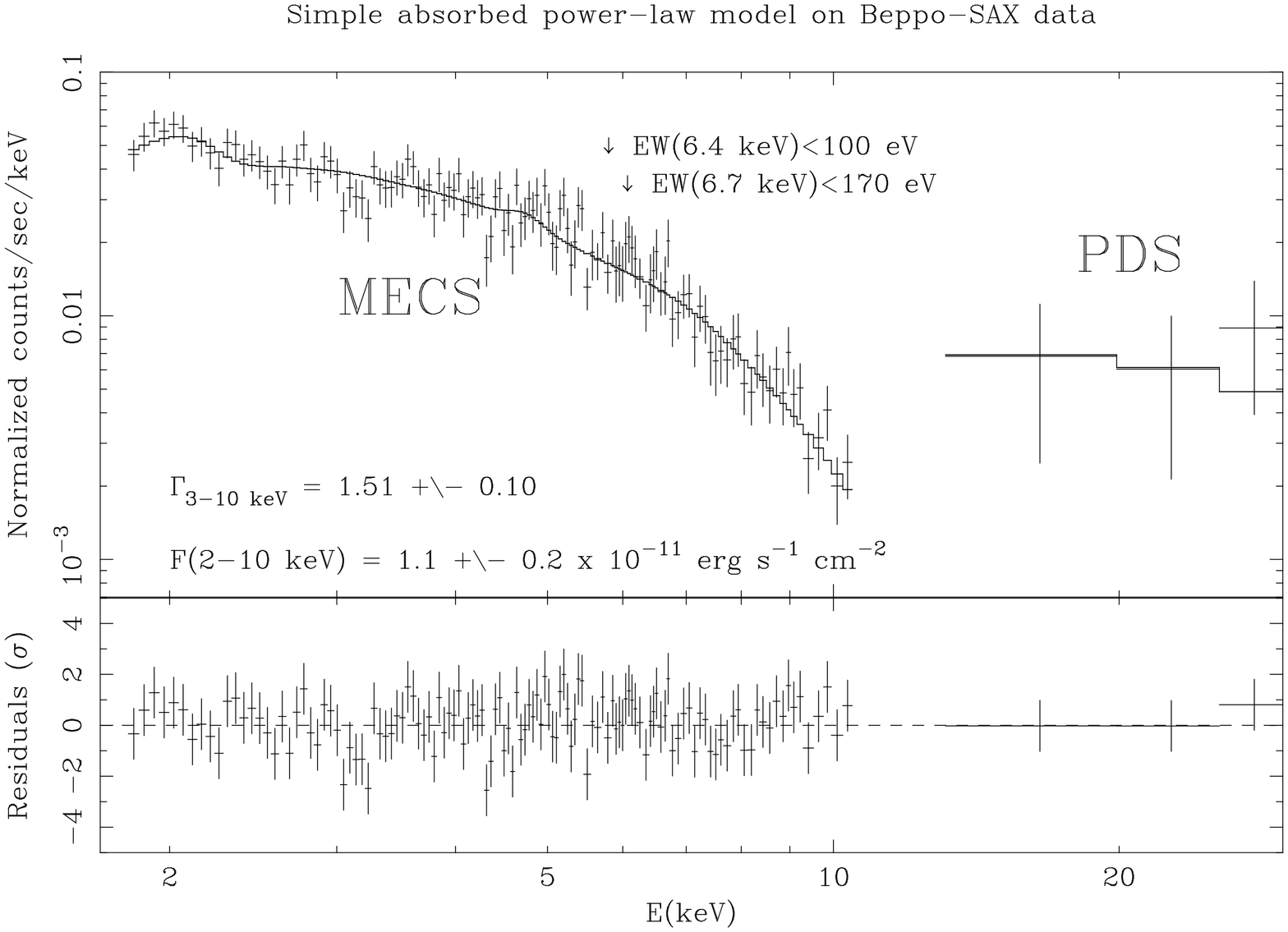,width=7.5cm,angle=-90}}
\caption[]{Spectra ({\it upper panel}) and residuals in units
of standard deviations ({\it lower panel}) when a power-law
model with Galactic ({\it i.e.} $N_H = 2 \times 10^{20} \ cm^{-2}$)
absorption is fitted to the MECS/PDS data.
The quoted photon spectral index and flux refer to this model.
90\% Upper limits on the
EWs of fluorescent iron lines are shown for the neutral
and ionized cases.}
\label{fig1}
\end{figure}

\begin{figure}
\centerline{\psfig{figure=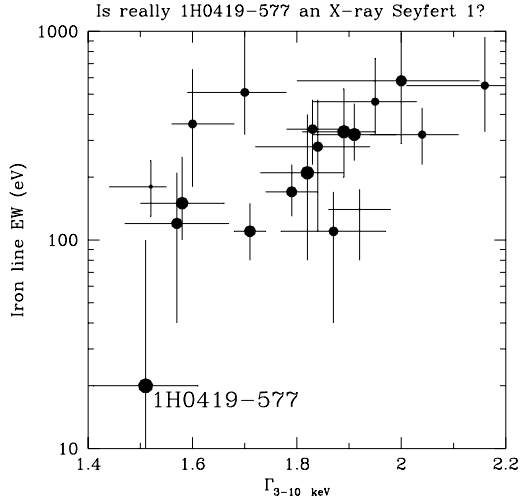,width=7.0cm,angle=0}}
\caption[]{Iron line EW vs. $\Gamma$ plot for the Seyfert 1 sample of [9].
The position of 1H0419-577 according to the Beppo-SAX
observation is indicated. Dot size is proportional to 2--10~keV luminosity.}
\label{fig2}
\end{figure}

The
1.8--45~keV
spectrum is well represented by a flat (photon index
$\Gamma_{3-10 \ keV} \simeq 1.5$) power--law
with photo-electric Galactic absorption
($N_H = 2 \times 10^{20} \ cm^{-2}$\cite{dl}, $\chi^2=130.6/153$ d.o.f., see Fig.~\ref{fig1}).
A spectral steepening is marginally
required for $E \simlt 3 \ keV$
($\Gamma_{soft} - \Gamma_{hard} \sim 0.3$):
$\Delta \chi^2$ for a broken (double) power--law = 2.6 (4.1) in
comparison to the simple power-law.
An iron line is not required either, the 90\% confidence level upper
limit on the equivalent width ($EW$) of a 6.4 (6.7)~keV
narrow ({\it i.e.} intrinsic Gaussian dispersion held fixed to 0)
line being 100 (170)~eV.
Neither a Compton reflection component nor broadening of the iron line
are statistically required by the data.
A summary of the best--fit parameters is shown in Table~\ref{tab1}.
\begin{table*}[hbt]
% space before first and after last column: 1.5pc
% space between columns: 3.0pc (twice the above)
\setlength{\tabcolsep}{1.5pc}
% -----------------------------------------------------
% adapted from TeX book, p. 241
\newlength{\digitwidth} \settowidth{\digitwidth}{\rm 0}
\catcode`?=\active \def?{\kern\digitwidth}
% -----------------------------------------------------
\caption{Beppo-SAX observation best-fit parameters}
\label{tab1}
\begin{footnotesize}
\begin{tabular}{lccccc} \hline \hline
\multicolumn{6}{c}{Panel a: power-law continuum modeling} \\ \hline
Model & $\Gamma_{hard}$ & $\Gamma_{soft}$ & $E_{break}$ & $N_{soft}/N_{hard}$ & $\chi^2$/d.o.f. \\
& & & (keV) & & \\ \hline
PO & $1.61\pm0.06$ & ... & ... & ... &130.6/153 \\
BKNPO & $1.56\pm0.10$ & $1.8\pm0.4$ & $3.0\pm1.0$ & ... & 128.0/151 \\
PO+PO & $1.55\pm0.09$ & $7^{\ddag}$ & ... & 3.7$^{\ddag}$ & 126.5/150 \\ \hline
\multicolumn{6}{c}{Panel b: iron line emission} \\ \hline
Model & $\Gamma_{hard}$ & $E$ & $\sigma$ & $EW$ & $\chi^2$/d.o.f. \\
& & (keV) & (keV) & (eV) & \\ \hline
PO+GA & $1.64^{+0.09}_{-0.08}$ & $6.2^{\ddag}$ & $0.7^{\ddag}$ & $180^{\ddag}$ & 128.1/151 \\
Narrow neutral line & $1.61^{+0.07}_{-0.06}$ & 6.4$^{\dag}$ & 0$^{\dag}$ & $17^{+83}_{-17}$ & 130.6/152 \\
Narrow ionized line & $1.63^{+0.06}_{-0.07}$ & 6.7$^{\dag}$ & 0$^{\dag}$ & $70^{+100}_{-90}$ & 129.1/152 \\ \hline \hline
\end{tabular}
\\
\noindent
$^{\ddag}$unconstrained \\
$^{\dag}$fixed \\
Photoeletric absorption from cold
material with $N_H=N_{H_{Gal}}=2 \times 10^{20} \ cm^{-2}$ was added to all the above models. \\
PO=power-law, BKNPO=broken power-law, GA=Gaussian line, PEXRAV=power-law+Compton reflection
\end{footnotesize}
\end{table*}
Such properties are rather
extreme among the Seyfert 1 galaxies observed in X--rays
(Fig.~\ref{fig2}).

\begin{figure}
\centerline{\psfig{figure=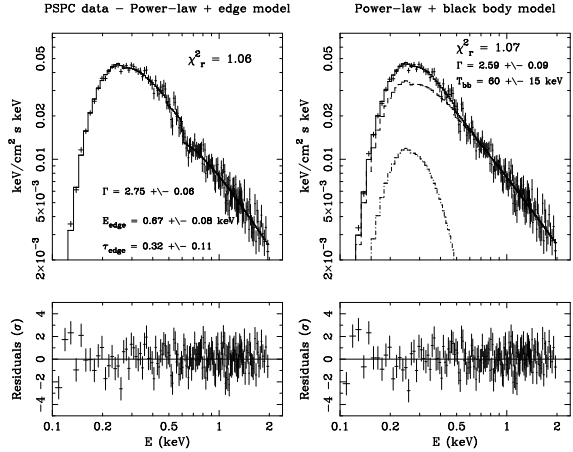,width=8.5cm,angle=-90}}
\caption[]{Unfolded ROSAT/PSPC spectrum ({\it upper panel}) and
residuals in units of standard deviations ({\it lower panel})
for the best-fit models. A much steeper ($\Gamma_{soft} = 2.6 \div 2.8$)
power-law component is required than in the MECS observations.}
\label{fig3}
\end{figure}

These results are compared with those obtained from a
ROSAT/PSPC observation.
of the same target, performed more than 4 years earlier
(see Table~\ref{tab2}).
\begin{table*}[hbt]
% space before first and after last column: 1.5pc
% space between columns: 3.0pc (twice the above)
\setlength{\tabcolsep}{1.5pc}
% -----------------------------------------------------
% adapted from TeX book, p. 241
% -----------------------------------------------------
\caption{ROSAT observation best--fit parameters}
\label{tab2}
\begin{footnotesize}
\begin{tabular}{lccccccc} \hline \hline
Model & $\Gamma_{soft}$ & $E$ & $EW$ or $\tau$ & $\Gamma_{ss}$ & $E_{break}$ or $T$ & $\chi^2/$d.o.f. \\
& & (keV) & (eV)/... & & (eV) & \\ \hline
BB & ... & ... & ... & ... & $101\pm1$ & 2901/151 \\
DISKBB & ... & ... & ... & ... & $142\pm1$ & 1916/153 \\
PO & $2.77^{+0.03}_{-0.02}$ & ... & ... & ... & ... & 172.7/151 \\
ED*PO & $2.75\pm0.06$ & $0.67\pm0.08$ & $0.32^{+0.08}_{-0.14}$ & ... & ... & 157.2/149 \\
PO+GA & $2.83^{+0.05}_{-0.07}$ & $1.56^{+0.23}_{-0.13}$ & $80\pm40$ & ... & ... & 162.7/149 \\
BKNPO & $2.53^{+0.14}_{-0.24}$ & ... & ... & $2.92^{+0.16}_{-0.11}$ & $800\pm200$ & 158.0/149 \\
PO+BB & $2.59\pm0.09$ & ... & ... & ... & $60^{+10}_{-15}$ & 159.2/149 \\ \hline \hline
\end{tabular}
PO=power-law,
ED=absorption edge, GA=Gaussian line, BKNPO=broken power-law,
BB=blackbody. d.o.f. = degrees of freedom
\end{footnotesize}
\end{table*}
The best--fit model requires a much steeper
power--law component
($\Gamma_{PSPC} - \Gamma_{MECS} \sim 0.7 \div 1.3$, see Fig.~\ref{fig3}).
We stress that the PSPC spectral index is inconsistent with that
measured by the MECS even if a broken power-law is used to fit the
latter data. The maximum amplitude of the systemic error on the
spectral index due to ROSAT/PSPC calibration ($\sim 0.2$ according
to \cite{fem} and \cite{t}) cannot account for the entire observed
difference.

\begin{figure}
\centerline{\psfig{figure=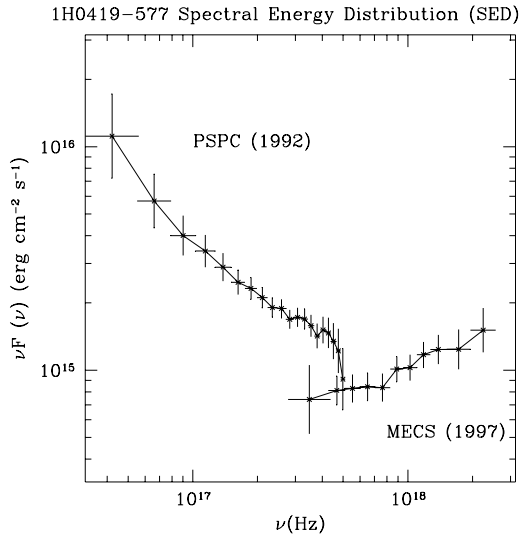,width=7.5cm,angle=0}}
\caption[]{1H0419-577 SED spectral shape
according to the 1992 ROSAT/PSPC and 1996 Beppo-SAX/MECS
observations}
\label{fig4} 
\end{figure}

The explanation of such a difference in terms of a soft
excess has been investigated; if
a power-law component is assumed to be present in the PSPC spectrum,
with $\Gamma_{hard} \equiv 1.55$ and the normalization is
free to vary, the data
require: a) either a very steep soft power-law component
($\Gamma_{soft} \simeq 2.9$), 
without any variation of the medium/hard X--ray spectrum;
b) or a broad ({\it i.e.}
multi-temperature) thermal component superimposed on the hard power-law,
whose normalization decreased by a factor
$\sim 2 \div 3$. However, these models are statistically worse at
$\ge 99.997\%$ level
than models
where the medium/hard X--ray photon index is allowed to vary;
moreover a 1995 ASCA observation of the same target
did not detect any strong soft excess below energies $E \simeq 1 \ keV$,
despite the fact that the 2--10~keV spectrum is similarly flat as
that measured by Beppo-SAX
(Marshall {\it et al.}, 1998, in preparation).
Although the hypothesis of a variable soft excess
cannot be completely ruled out, the above results suggest
that the observed difference
between 1992 PSPC and 1996 Beppo-SAX spectra is due to a ``true'' change
of the Comptonized medium/hard X--ray spectrum
(see also the Spectral Energy Distribution in Fig.~\ref{fig4}).
Further monitoring of the broadband X--ray spectrum of this source is
needed to clarify this issue. 

\section{Variable soft X--ray spectrum: a chance to test Comptonization scenarios}

$\Delta \Gamma \sim 1$ could occur: a) in
a two-phase disk-corona accretion disk model \cite{hm}, if the system undergoes a transition between
a scattering optically thick ($\tau \sim 1$) and thin ($\tau \simlt 0.1$) regimes \cite{hmg}, provided
the plasma is far from being pair-dominated; b) in a viscous-less two-phase shock accretion disk
model \cite{etc}),
as the effect of a decrease of the accretion rate from quasi-- to sub--Eddington regimes.
The latter case would suggest a link between the radio--quiet AGN
and the Black Hole Candidates, which are well known to display transitions between soft/high
(typical energy photon indices $\alpha \simeq 1.5$) and hard/low states
($\alpha \simeq 0.3 \div 0.9$). The above data however do not resemble the
archetypical example of such a connection, RE1034+39 \cite{pdo}.
In this source the ROSAT/PSPC spectrum could be interpreted in terms
of multi-temperature blackbody, suggesting that the soft state
represent the occurrence of quasi-Eddington accretion rate events, in which the
bulk of the emission comes from the energy release due to viscous dissipation in the disk.
However, this is not the case for 1H0419-577 (cfr. the
first two rows in Tab.~\ref{tab2}).
A change of the Comptonized spectrum provides a natural explanation for the relative faintness
of the iron line as well, which could suffer a delayed response to the - currently still unknown -
variability pattern of the underlying continuum.

\section*{Acknowledgments}

Discussions with Dr. F.Haardt about the disk-corona models significantly
improved
the quality of this paper.

\end{document}